\documentstyle[12pt]{article}
\newcommand{\be}{\begin{equation}}
\newcommand{\ee}{\end{equation}}
\newcommand{\ba}{\begin{eqnarray}}
\newcommand{\ea}{\end{eqnarray}}
\newcommand{\baa}{\begin{eqnarray*}}
\newcommand{\eaa}{\end{eqnarray*}}
\newcommand{\bb}{\begin{thebibliography}}
\newcommand{\eb}{\end{thebibliography}}

\newcommand{\ci}[1]{\cite{#1}}
\newcommand{\bi}[1]{\bibitem{#1}}
\begin {document}

\begin{center}
{\Large \bf Comment on the Burkhardt -- Cottingham and Generalized
Gerasimov--Drell--Hearn Sum Rules for the neutron}\\[1cm]
{J. Soffer}\\[0.3cm]
{\it Centre de Physique Th\'eorique - CNRS - Luminy,\\
Case 907 F-13288 Marseille Cedex 9 - France}
\footnote{E-mail: soffer@frcpn11.in2p3.fr}
\\[0.3cm]
and\\[0.3cm]
{O.V. Teryaev}\\[0.3cm]
{\it Bogoliubov Laboratory of Theoretical Physics\\
Joint Institute for Nuclear Research, Dubna\\
Head Post Office, P.O. Box 79, 101000 Moscow, Russia}
\footnote{E--mail: teryaev@thsun1.jinr.dubna.su}\\[1.2cm]
\begin{abstract}
A description of the generalized Gerasimov--Drell--Hearn
sum rule for neutron is suggested, using
its relation to the
Burkhardt--Cottingham sum rule.
\end{abstract}
\end{center}

PACS numbers: 11.50.Li, 13.60.Fz, 13.88.+e

\newpage

The generalized Gerasimov-Drell-Hearn (GDH) sum rule \ci{Ger,DH} is
just being tested experimentally \cite{E143} and
the proton data are in the good agreement with our prediction
\cite{ST93,ST95}, making use of its relation to the Burkhardt-Cottingham
sum rule. They also agree with a new estimate of the
contributions from low-lying resonances \cite{Ioffe}.
We should stress, that such a similarity is by no means surprising,
since the dominant magnetic form factor of $\Delta(1232)$ is
contributing entirely through the structure function $g_2$ \cite{ST95},
which is the key ingredient of our approach.
Given the fact that the experimental data \cite{E143} are also available
for a neutron target (although with a poor accuracy),
we present here the quantitative prediction
for the neutron case.

Let us consider the $Q^2$-dependent integral
\be
I_1(Q^2)={2 M^2\over {Q^2}} \int^1_0 g_1(x,Q^2) dx.
\label{I1}
\ee
It is defined for {\it all} $Q^2$, and $g_1$ is the obvious
generalization for all $Q^2$ of the standard scale--invariant $g_1(x)$.
Note that the elastic contribution at $x=1$ is not
included in the above sum rule. One recovers then at $Q^2=0$
the GDH sum rule
\be
I_1(0)=-{\mu_A^2 \over 4},
\ee
where $\mu_A$ is the nucleon anomalous magnetic moment in nuclear magnetons.
While the $I_1(0)$ is always negative, its value at large $Q^2$ is determined
by the $Q^2$-independent integral $\int^1_0 g_1(x) dx$,which is
positive for proton and negative for neutron.

It is possible to decompose $I_1$ as the difference of the contributions
from two form factors $I_{1+2}$ and $I_2$
\be
I_1=I_{1+2}-I_2,
\ee
where
\be
I_{1+2}(Q^2)={2 M^2\over {Q^2}} \int^1_0 g_{1+2}(x) dx,
\;\;\;\;I_2(Q^2)={2 M^2\over {Q^2}} \int^1_0 g_2(x) dx,
\ee
\be
g_{1+2}=g_1 + g_2.
\ee
There are solid theoretical arguments to expect a strong $Q^2$-dependence
of $I_2$. It is the well-known Burkhardt-Cottingham(BC) rule \ci{BC},
derived independently by Schwinger \ci{Sch},
using a rather different method. It states that
\be
I_2(Q^2)={1\over 4}\mu G_M (Q^2)[\mu G_M (Q^2) - G_E (Q^2)],
\ee
where $\mu$ is the nucleon magnetic moment, $G$'s denoting the familiar
Sachs form factors which are dimensionless and normalised to unity
at $Q^2=0$. For large $Q^2$ one can neglect the r.h.s. and get
\be
\int^1_0 g_2(x) dx=0.
\ee

In particular,
\be
I_2(0)={\mu_A^2+\mu_A e \over 4},
\ee
$e$ being
the nucleon charge in elementary units.
To reproduce the GDH value one should have
\be
I_{1+2}(0)={\mu_A e \over 4}.
\ee
Note that $I_{1+2}$ does not differ from $I_1$ for large $Q^2$ due to the
BC sum rule, but it is {\it positive} in the proton case.
It is possible
to obtain a smooth interpolation for  $I^p_{1+2}(Q^2)$
between large $Q^2$ and $Q^2=0$ \ci{ST93},namely

\be
I^p_{1+2}(Q^2)=\theta(Q^2_0-Q^2)({\mu_A \over 4}- {2 M^2 Q^2\over
{(Q^2_0)^2}} \Gamma^p_1)+\theta(Q^2-Q^2_0) {2 M^2\over {Q^2}}
\Gamma^p_1,
\ee
where $\Gamma^p_1=\int^1_0 g^p_1(x) dx$.
The continuity of the function and of its derivative is guaranteed
with the choice $Q^2_0=(16M^2/\mu_A) \Gamma^p_1 \sim 1GeV^2$,
where the integral is given by the world average proton data.
It is quite reasonable to distinguish
the perturbative and the non-perturbative regions. As a
result one obtains a crossing point at $Q^2 \sim 0.2 GeV^2$, below the
resonance region \ci{ST93}, while the positive value at $Q^2=0.5 GeV^2$
is in a good agreement with the E143 data \ci{E143}.

To generalize this approach to the neutron case, one needs a
similar parametrization for the neutron. Since the value at
$Q^2=0$ is equal to zero, it is not sufficient to limit oneself
to the simplest linear parametrization and one needs to
add two terms, one quadratic and one cubic in $Q^2$. A simple
parametrization
providing the continuity of the function and its first and second
derivative is

\be
I^n_{1+2}(Q^2)={2 M^2 \over
{Q_0^2}} \Gamma^n_1[{ Q^2 \over {Q_0^2}}\theta(Q^2_0-Q^2)
[6-8 {Q^2\over { Q_0^2}}+ 3 ({Q^2 \over { Q_0^2}})^2]
+\theta(Q^2-Q^2_0){Q_0^2 \over {Q^2}}]
\ee

Note that because of the extra terms in the parametrization,
$Q_0^2$ is not determined by the continuity
conditions, and we are taking the same value as in the proton case.
The parametrization seems to be rather natural in terms of the
$Q^2$-dependent integral $\Gamma^n_{1+2} (Q^2)$,which
is equal to its asymptotic value $\Gamma^n_{1+2} (\infty)=\Gamma^n_1$
down to $Q_0^2$. Below $Q_0^2$ it is just

\be
\Gamma^n_{1+2}(Q^2)= ({ Q^2 \over {Q_0^2}})^2 \Gamma^n_1
[6- 8{Q^2\over { Q_0^2}}+ 3 ({Q^2 \over { Q_0^2}})^2].
\ee

To parametrize the elastic contribution to the BC sum rule we need
the neutron elastic form factors. While the electric one might be
neglected, the magnetic form factor
is well decribed by the dipole formula \cite{FF}

\be
G_M (Q^2)={1 \over {(1+Q^2/0.71)^2}}.
\ee

The plot representing $\Gamma^n_1 (Q^2)$ is displayed
on Fig.1, where we used $\Gamma^n_1=-0.04$.  One can see a behaviour
distinct to the case of
the resonance model \cite{Ioffe}, since the minimum occurs at $-0.06$
for $Q^2=0.7 GeV^2$. In addition it is worth noting that our prediction
is typically below the QCD extrapolation, since we subtract $I_2$
(which is a sharp positive contribution both for proton and neutron)
from smooth $I_{1+2}$.
In the resonance model it is above the QCD extrapolation for the neutron
and below for the proton.
We hope it will be possible to distinguish between these two predictions,
as soon as more accurate data will be available from CEBAF.

The success of our model may be considered as an indirect check of the
BC sum rule, but it could be also due to a possible cancellation \cite{ST95}
of non-scaling Regge cuts \cite{IoLi} in $I_1$, in such a way to preserve
our predictions.

We are indebted to Keith Griffioen and Sebastian K\"uhn
for an interesting discussion
which led us to write this short paper.

\bb{99}
\bi{Ger} S.B.Gerasimov, Yad. Fiz. {\bf 2}, 598(1965)
[Sov. J. Nucl Phys. {\bf 2}, 430(1966)].
\bi{DH} S.D.Drell and A.C.Hearn, Phys. Rev. Lett. {\bf 16}, 908(1966).
\bi{E143} E143 Collaboration, K.Abe et al.,Phys.Rev.Lett. {\bf78}, 815
(1997).
\bi{ST93} J.Soffer and O.Teryaev, Phys. Rev. Lett. {\bf 70}, 3373(1993).
\bi{ST95} J.Soffer and O.Teryaev, Phys. Rev. {\bf 51}, 25(1995).
\bi{Ioffe} V.D.Burkert and B.L.Ioffe, Phys.Lett. {\bf B296}, 223(1992).
JETP {\bf 105}, 1153(1994).
\bi{BC} H.Burkhardt and W.N.Cottingham, Ann. Phys. (N.Y.) {\bf 16}, 543(1970).
\bi{Sch} J.Schwinger, Proc. Nat. Acad. Sci. U.S.A. {\bf 72}, 1559(1975).
\bi{FF}  A.Lung et al., Phys. Rev. Lett. {\bf 70}, 718(1993).
\bi{IoLi} B.L.Ioffe, V.A.Khoze, L.N.Lipatov, {\it Hard Processes},
(North-Holland,Amsterdam, 1984).
\eb

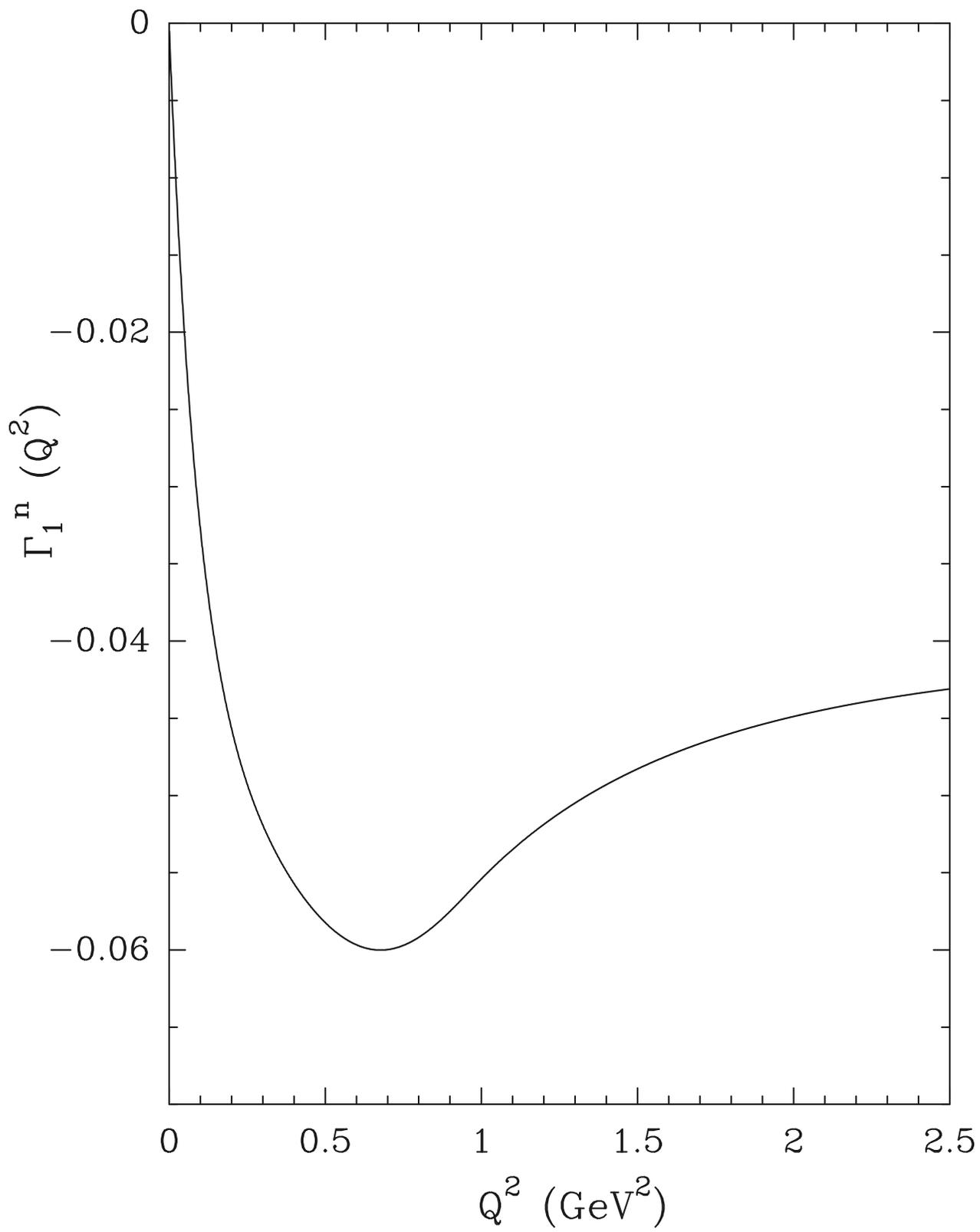
\end{document}